\documentclass[twocolumn,superscriptaddress,amsmath,amssymb,prl]{revtex4}

\usepackage{dcolumn}
\begin{document}

\title{Infrared Lorentz Violation and Slowly Instantaneous Electricity}
\author{Gia Dvali}
\affiliation{Center for Cosmology and Particle Physics, Department of Physics, New York University, NY 10003}
\email{gd23@nyu.edu}
\author{Michele Papucci}
\email{papucci@berkeley.edu}
\affiliation{Department of Physics, University of California, Berkeley, CA 94720 \\
 and Lawrence Berkeley National Laboratory, Berkeley, CA 94720}
\author{Matthew D. Schwartz$^2$}
\email{mdschwartz@lbl.gov}

%\date{\today}% It is always \today, today, but you may specify any date with \date.

\begin{abstract}
We study a  modification of electromagnetism which violates Lorentz invariance at large
distances. In this theory, electromagnetic waves are massive, but
the static force between charged particles is Coulomb not Yukawa. At very short distances
the theory looks just like QED. But for distances larger than $1/m$ the massive dispersion relation
of the waves can be appreciated, and the Coulomb force can be used to communicate faster than the
speed of light. In fact, electrical signals are transmitted instantly, but take a time $\sim 1/m$ 
to build up to full strength. 
After that, undamped oscillations of the electric field are set in
and continue until they are dispersed by the arrival of the Lorentz-obeying part of the transmission.
We study experimental constraints on such a theory and find that the Compton wavelength of the photon
may be as small as 6000 km. This bound is weaker than for a Lorentz-invariant mass, 
essentially because in our case the Coulomb constraint is removed.
\end{abstract}

%\pacs{}
\maketitle

Quantum Electrodynamics and Lorentz invariance are firmly established to
astounding accuracy.
But most of the evidence for both of these ideas comes from
short distance tests. 
These tests are justified, because we expect new ideas
to become relevant at short distances, but there are plenty of signs
that something strange may happen at large distances too (for example, dark energy),
where the constraints are much weaker.
In fact, even drastic modifications of the standard model may not be ruled out.
To explore this possibility,
we consider a scenario in which the photon has a
Lorentz-violating mass $m$. We find a number of counterintuitive results. The
photon mass does not slow the field down, but speeds it up. In fact, the
electric field can now transmit signals instantly, although it takes a
fixed time $\sim 1/m$ for the signal to mature. We also find that the bound on
the photon mass is actually weaker than for the Lorentz invariant case.
% This
%is essentially because Coulomb's law is unmodified.

%\section{Fast Electricity}
The simplest infrared Lorentz-violating modification of an Abelian gauge
theory is the addition of rotationally invariant mass terms.
(During the preparation of this work, we learned about an independent work 
\cite{Gabadadze:2004iv}, which introduces a similar system.)
The most general possibility is
\begin{equation}
  \mathcal{L_{\mathrm{gen}}} = - \frac{1}{4} F_{\mu \nu}^2 + \frac{1}{2} m_0^2
  A_0^2 - \frac{1}{2} m^2 A_i^2 + A_{\mu} J^{\mu} .
\end{equation}
For $m_0 = m$ this is the Lorentz-invariant Proca theory~\cite{Proca}. The
equations of motion imply the generalized Proca constraint $m_0^2 \partial_t A_0
= m^2 \partial_i A_i$ and for the static case let us immediately solve for the
scalar potential
\begin{equation}
  A_0 = \frac{1}{\triangle - m_0^2} J_0  . \label{scalpot}
\end{equation}
Thus, for $m_0 \neq 0$ the static electrical force is Yukawa. That is,
except for the special case $m=0$, when the theory degenerates to
the massless one~\cite{footnote1}.

The main subject of this paper is a different special case, $m_0 = 0$, for which
the Lagrangian reads
\begin{equation}
  \mathcal{L} = - \frac{1}{4} F_{\mu \nu}^2 - \frac{1}{2} m^2 A_i^2 + A_{\mu}  J^{\mu} .
\label{ourlag}
\end{equation}
Using current conservation, $\partial_{\mu} J^{\mu} = 0$, the equations of
motion can be reduced to
\begin{equation}
  A_i =- \frac{ \delta_{ij}-\frac{\partial_i \partial_j}{\triangle}}
{\Box    +  m^2} J_i 
\hspace{2em} A_0 = \frac{J_0}{\triangle} . \label{aeqs}
\end{equation}
These seem to imply that electromagnetic waves are transverse but massive,
while the force between charges is Coulomb and instantaneous. However, for $m
= 0$ these equations are precisely those of massless electrodynamics in the
Coulomb gauge, where we know the Coulomb force is not instantaneous, but
causality is obscure.

To clarify the situation, it is helpful to work directly
with the field strengths, which are the only degrees of freedom that couple
to matter.
From \eqref{aeqs} follow the wave equations for the electric and magnetic fields:
\begin{eqnarray}
  ( \square + m^2 ) \vec{B} &=& \vec{\nabla} \times {\vec{J}} \\
  ( \square + m^2 ) E_i &=& \partial_i J_0 - \partial_t J_i - \frac{m^2}{\triangle}
  \partial_i J_0 \label{efield}
\end{eqnarray}
where $\vec{B} =- \vec{\nabla} \times \vec{A} $ and $E_i = \partial_t A_i - \partial_i A_0$.
To see the Coulomb force, take a static situation where $\partial_t = 0$. 
Then $\square = -\triangle $ and the equation of motion for the electric field is the same as
in the static massless case: $\triangle E_i = - \partial_i J_0$. This implies
that there is a long distance Coulomb force in our theory, but tells us
nothing about its dynamics.

To understand the dynamics, it is helpful to separate the electric field into
two parts:
\begin{eqnarray}
  E_i &=& E_i^{\mathrm{proca}} + \partial_i \Phi \label{esplit}\\
  ( \square + m^2 ) E_i^{\mathrm{proca}} &=& \partial_i J_0 - \partial_t J_i \\
  ( \square + m^2 ) \Phi &=& - \frac{m^2}{\triangle} J_0 \label{instant}
\end{eqnarray}
Now we see that $E^{\mathrm{proca}}_i$ and $\vec{B}$ are identical to the electric and
magnetic fields in the Proca theory: transverse electromagnetic waves are
massive, and the potential they induce between charges has Yukawa's form
$\frac{1}{r} e^{- m r}$. All the new effects are therefore contained in the
new scalar potential $\Phi$, which propagates like a massive field and has an
effective source $\rho_{\mathrm{eff}} = - \frac{m^2}{\triangle} J_0$. For a single
point charge at the origin, $J_0 =4 \pi q \delta^3 ( r )$ and so
$\rho_{\mathrm{eff}} =  q \frac{m^2}{r}$. Then the solution to \eqref{instant}, in the
static case, is
\begin{equation}
  ( -\triangle + m^2 ) \Phi =  q \frac{m^2}{r} \quad \Rightarrow \quad \Phi =
  \frac{q}{r} - \frac{q}{r} e^{- m r} \label{plaspot} .
\end{equation}
The Yukawa part of this potential precisely cancels the Yukawa force mediated
by $E^{\mathrm{proca}}_i$, leaving the Coulomb force which confirms the calculation
directly from \eqref{efield}. So the cartoon we have is of regular massive electric and
magnetic fields enhanced by another electric field induced by a spurious
plasma which has a charge distribution $\rho_{\mathrm{eff}} = q \frac{m^2}{r}$
around each source.

Now suppose we wiggle the source, how long does it take for the Coulomb field
at a distance $r$ to be affected? 
We shall now show that the system exhibits the following behavior. 
First, for a localized source  
%$J_0(r,t)$
the response is ``slowly-instantaneous''. That is,
%meaning that 
the electric field at arbitrarily large distances
$r \gg m^{-1}$ starts responding immediately to the time variation of a source at
$r=0$, but the amplitude of the response takes time $t \sim m^{-1}$ to build up to full strength. 
Second, the time duration of the response can be arbitrarily longer than  
the time-dependence of the source.
% $J_0(r,t)$,  $\Delta t$.  

To demonstrate these effects, it is enough to consider the
contribution to the electric field from $\Phi$, which mediates all the Lorentz violation.
We suppose the source $J_0$ is localized and moves for only a finite time interval:
$\partial_t J_0\, \neq 0$ only for $0 \leqslant t \leqslant \Delta t$.  
In such a case, a good order
parameter for analyzing the retardation properties of the system is the time derivative 
$\dot{E}^{\Phi}_j\, =\, \partial_j\dot \Phi$.

From (\ref{instant}), by  performing a Fourier transformation in the spatial directions
but keeping time-dependence explicit, we arrive at
%get the following
%equation for the
%Fourier 
%image of $\dot{E}^{\Phi}_j$
 \begin{equation}
\label{psi}
\partial_t^2 \tilde{\dot{E}}^\Phi_j (p,t)\, + \, (p^2 \, + \, m^2)\tilde{\dot{E}}^\Phi_j(p,t) \, 
= -i {p_j \over p^2} \, m^2 \dot{\tilde{J_0}}(p,t)
\end{equation}
where $p = |\vec{p}|$. 
For each momentum mode, 
this is the equation of a harmonic oscillator subjected to an external force of the duration 
$\Delta t$.
%The above two features are now clear.  
The oscillator will start responding immediately to the force, but it takes a time $\sim m^{-1}$ 
for the response to develop.  After the external force is switched off,
the oscillator generically  will be left excited 
and will continue to oscillate.

It is instructive to demonstrate the above effects by solving an explicit example of a physical situation.
Suppose we have two opposite charges at the origin, and separate them into a dipole of strength $\mu$
in the $z$ direction at time $t=0$.
(As far as $\dot E$ is concerned, this is the same
moving a charge $q$ a distance $\Delta L = \mu /q$.)
The current is
\begin{eqnarray}
J_0 &=& - 4 \pi \mu \partial_z \delta^3(r) \Theta(t)\\
J_z &=& - 4 \pi \mu \delta^3(r)\delta(t),\quad J_x = J_y=0
\end{eqnarray}
Then $\Phi$ satisfies
\begin{equation}
  ( \partial_t^2 - \triangle + m^2 ) \Phi = - \mu \partial_z \frac{m^2}{r} \Theta ( t ) \label{source}
\end{equation}
Instead of solving for $\Phi$ exactly, we will pull of the important part of
the solution
\begin{eqnarray}
  \Phi &=& \Phi_a + \Phi_b\\
  \Phi_a &=& \mu \partial_z \frac{1}{r} \left[ 1 - \cos ( m t ) \right] \Theta ( t )
\end{eqnarray}
which leaves
\begin{equation}
   ( \partial_t^2 - \nabla^2 + m^2 ) \Phi_b = - 4 \pi
\mu \partial_z  \delta^3 ( r ) [ 1 - \cos (  m t )] \Theta ( t )
\end{equation}
The point of this is that now the source for $\Phi_b$ is localized, at $r =
0$, and so $\Phi_b$ will vanish outside the light cone: $\Phi_b=0$ for $t > r$. Thus the
Lorentz-violating effects are contained entirely $\Phi_a$, for which we know the complete solution.
So we get for the
order parameter $\dot{E}_j$
\begin{equation}
\dot{E}_j = \mu m \partial_j \partial_z \frac{1}{r} \sin(m t) \Theta(t), \quad r>t \label{exsol} .
\end{equation}
This is an exact result. The electric field outside the light-cone 
is that of a dipole in massless electrodynamics, but
is modulated by undamped oscillations of frequency $\omega = m$.

After time $t=r$, when the Lorentz-invariant signal arrives, we do not have a simple closed form
expression for the electric field. But it is not hard to show
that the oscillations will become damped, matching smoothly
onto the static solution for $t \gg r$. An exact expression for $\dot \Phi$ can be derived by
integrating the retarded Green's function for a massive scalar against the non-local source
in \eqref{source}. We reserve for the reader the pleasure of demonstrating that in this case
\begin{eqnarray}
\dot{E}^{\Phi}_j &=&\mu m^2 \partial_j \partial_z\frac{ t}{r}
 \int_0^{r/t} dx {\cal J}_0 (m t \sqrt{1-x^2} ),
\quad t>r \nonumber\\
&\approx& \mu m^3 \partial_z \partial_j \frac{ r^2}{t} {\cal J}_1(m t) + \cdots, \quad t \gg r
\end{eqnarray}
The Bessel function ${\cal J}_1(m t)$ dies as $ (m t)^{-1/2}$ for large $t$.  Similar large time
behavior holds for the contribution to $\dot E$ from the Proca field in \eqref{esplit}
and we conclude that all the oscillations eventually die off.

Returning to \eqref{exsol},
 we can observe all the essential properties of the slowly-instantaneous 
signal propagation in our system: 
at arbitrarily large distances, $r\gg m^{-1}$, the full response sets in only after the time
$m^{-1}$;
and the oscillations continue for arbitrarily long times,
until the point $r$ enters the light cone of the source.

The appearance of the oscillating electric field is not so unexpected.  
Because the space-like photons are massive bosons
 the homogeneous 
 equation for the electric field \eqref{efield} has 
 free massive wave solutions with arbitrary momenta, 
 and in particular an oscillating solution 
\begin{equation}
\label{osce}
E_j \, = \, a_j \cos (mt)\, + \, b_j \sin (mt).
\end{equation} 
These oscillations are not any different from the coherent oscillations
of any other massive boson field, 
and describe a state with non-zero occupation number of zero momentum photons.

With this understanding we can proceed to study constraints on the model, to 
see what the experimental bound is on $m$. 
First, observe that the mass term $m^2 A_j^2$ contributes
to the Hamiltonian, and will generically lead to extremely strong constraints, from
limits on the known energy density of the universe. For example, 
the contribution from the vector potential of the galactic magnetic field would force $m<10^{-27}$ eV. 
However,
it has been shown~\cite{Adelberger:2003qx} that in the Lorentz-invariant Proca case,
this constraint is removed completely if the photon mass is spontaneously generated, due
to the creation of vortices, like in a type II superconductor.
We will now review this argument and show that it holds for 
a Lorentz-violating mass as well.

The mass in \eqref{ourlag} can be generated spontaneously from a Lorentz-violating 
but gauge invariant Lagrangian
\begin{equation}
{\cal L} = -\frac{1}{4} F_{\mu \nu}^2 + |D_i H|^2 + V(H) + A_\mu J^\mu \label{higgs}
\end{equation}
where $D_i H = \partial_i H + q_H i  A_i H$ and $V$ is some potential for Higgs field $H$. If the vacuum has
$\langle H \rangle = v$, we are returned to \eqref{ourlag} at low energy, with $m \sim q_H v$. 
Note that the charge of the  Higgs, its mass, and the mass of the photon are all free parameters.

Equivalently, we can rewrite the mass term as
$m^2A_j^2 \rightarrow |H|^2 (q_H A_j \, + \, \partial_j\theta)^2$, where
$H$ is now the modulus field, and $\theta$ its Goldstone phase. By separating out $\theta$ as an independent field,
we see that it is free to assume a configuration which tries to cancel the contribution to the Proca energy
from $A_j$. Suppose there is a uniform magnetic field $B_z$, then we can take $A_\phi = B_z r$, where $(r,\phi,z)$ 
are cylindrical coordinates. 
It follows that a minimum energy configuration would have the Higgs winding
around the $z$ axis with $1/r \partial_\phi \theta =  q_H B_z r$. 
Of course, $\theta$ and $\phi$ are periodic, so we can only have vortex configurations, such as
$\theta = c \phi$ where $c \approx q_H B_z r^2$ is an integer. These vortices will form and 
distribute, resulting in a total vortex flux $n/q_H$, which should be close to $\sim B_z r^2$
for a region of size $r$. In particular, if there are very many vortices in the region with $r<1/m$, that is
if $n = q_H B/m^2 \gg 1$, then
effect of the mass on the magnetic field will be completely screened.

The vortex configurations are only possible if $H$ has isolated zeros at the vortex
cores, which is why this argument fails without the modulus scalar. The cores will have a size $\sim 1/m_H$,
giving a gradient energy $\sim m_H^2 v^2$. This should be smaller than
the Proca energy $\sim q_H^2 v^2 B^2_z r^2$ if the vortices are to be created at all. But
if the system satisfies $q_H B_z r/m_H \gg 1$ then the vortices can be created classically
at almost no cost. On the other hand, as $m_H \to \infty$, the vortices disappear,
which is, of course, consistent with decoupling the modulus.

The only difference between this case and one discussed in~\cite{Adelberger:2003qx}
is the absence of a time derivative for $H$ in the Lagrangian. 
But since the diffusion of the Proca energy and the screening of the magnetic field
is from static vortices, the argument goes through unchanged.
So we conclude that also in this case the constraint from the Proca energy of the universe can be ignored.
Moreover, all constraints derived from the decay of planetary or galactic magnetic fields,
become irrelevant if the vortices can proliferate. For example, on the earth $B \sim 10^{-2} {\mathrm eV}^2$,
and $r \sim 10^{13} {\mathrm eV}^{-1}$. For a photon mass of $10^{-13} eV$, we find $n=q_H B/m^2 = 10^{24} q_H$
and $q_H B_z r/m_H \sim 10^{11}{\mathrm eV} q_H/m_H$. We can easily arrange for both of
these numbers to be huge
because $q_H$ and $m_H$ are free parameters.  
Note that including the Higgs does not to take us out of the Proca phase, except in tiny
regions near the vortex cores;
waves are still massive and the slowly instantaneous behavior of electricity persists.

The next strongest constraints on the Proca theory come from precision tests of Coulomb's law. One
tries to fit the electric force to a form $r^{-2+\delta}$ and by bounding $\delta$, one bounds $m$.
In the Proca case, this gives $m < 5\times 10^{-15} {\mathrm eV}$~\cite{Tu:2005ge}.
However,
in our case, Coulomb's law is unchanged, and so these bounds simply do not apply. In fact, the
only experimental constraints which will apply are dynamical ones, involving
electromagnetic waves. For example, there is a constraint on the Proca theory coming
from 
the time delay dispersal in the arrival of light from distant optical 
pulses~\cite{Tu:2005ge}.
This puts a limit 
$m \lesssim 2 \times 10^{-13} \mathrm{eV} \sim ( 6000 \mathrm{km} )^{- 1} \sim (20 \mathrm{ms})^{-1}$.
%If the photon mass is near this upper bound, we could in principle detect the Lorentz-violating effects
%by looking for resonances in the Coulomb field at a distance $200$ km from a moving source.

It is worth briefly considering how the Lorentz violation in \eqref{ourlag} or \eqref{higgs} 
might come about. It is possible to formally restore Lorentz invariance to the Lagrangian, 
by introducing
a new vector field $B_\mu$ and replacing the $(D_iH)^2$ term with
\begin{equation}
{\cal L}_B =  D_\mu H(B^\mu B^\nu - \eta^{\mu \nu} B_\alpha B^\alpha)D_\nu H^\star
\end{equation}
We can think of a $B_\mu$ as a spurion with the fixed value $B_\mu = (1,0,0,0)$. Or $B_\mu$ can
be related to some cosmological vector which happens to have a time-like vacuum expectation value.
For example, if we write $B_\mu = \partial_\mu \phi$ then the term becomes
\begin{equation}
{\cal L}_\phi =  D_{\mu} H  [ (
  \partial^{\mu} \phi ) ( \partial^{\nu} \phi ) - \eta^{\mu \nu} (
  \partial_{\alpha} \phi ) ( \partial^{\alpha} \phi ) ] D_{\nu}
  H^{\star}  \label{phicond}
\end{equation}
Then if $\phi$  is some background scalar field with expectation value 
$\langle \phi \rangle \sim t$ we are  brought back to~\eqref{higgs}. Such a situation
can be achieved, for example, from ghost condensation~\cite{Arkani-Hamed:2003uy} in which
derivatives of $\phi$ are given a potential with non-zero minimum. But we may also think of $\phi$ as being
related to other background fields, such as the scale factor of the universe $a(t)$.
We should also mention that although the form of \eqref{phicond} looks unnatural, it
is in fact protected by the symmetry $H \to exp(i f(\phi)) H$, for an arbitrary function $f$.
In the vacuum, this symmetry and gauge invariance are almost completely broken, leaving only
$A_\mu \to A_\mu + \partial_\mu f(t)$, which is the residual
symmetry keeping $m_0=0$ in \eqref{ourlag}.

Next, we can consider making this type of Lorentz-violating modification to a gauge
theory with self-interactions. If we drop the Higgs, then 
for a non-Abelian theory, the biggest change is the appearance of a scale where
perturbation theory breaks down, $\Lambda \sim 4 \pi m / g$. This is the same scale as for a
Lorentz-invariant theory, and can be traced to the scattering of
longitudinal modes. In our case there are no propagating longitudinal modes,
but there are additional degrees of freedom $\Phi^a$ which are the physical
cause of the superluminal effects, and we must take their interactions
seriously. For QCD, there is no legitimate constraint from low energy, where
it {\it is} non-perturbative, but asymptotic freedom up to $\sim 100$ GeV
forces the gluon to be very heavy, which is ruled out, for example,
by the observation of jets. The strong coupling could be moderated by a Higgs, but this new
Higgs would have to be colored and it is unlikely that such a colored particle
could have avoided characterization until now.  For the weak interactions, the
gauge bosons are already massive. If we changed the mass term to the
Lorentz-violating form, tree level results would be unchanged, because the
longitudinal mode does not couple to matter, but 1-loop corrections would
contradict electroweak precision data. Moreover, the weak force would be
Coulomb, not Yukawa -- it would not be weak! So we conclude that the non-Abelian
gauge theories in the standard model cannot accommodate Lorentz-violating
masses. But there is no inconsistency in the non-Abelian generalization, and
Lorentz-violating extensions of the standard model may be worthwhile to
explore.

For gravity, the most general Lagrangian with Lorentz-violating mass terms for the graviton
is~\cite{Rubakov:2004eb}
\begin{eqnarray}
 && \mathcal{L} = gR + \\
&& m_0^2 h_{00} + 2 m_1^2 h_{0j}^2  - m_2^2 h_{i j}^2 + m_3^2 h_{i i}^2 - 2
  m_4^2 h_{00} h_{i i}  \nonumber
% \qquad\qquad m_4^2 = m_0^2 ( 3 m_3^2 - m_2^2 )
\end{eqnarray}
The numerical coefficients are chosen so that if $m_0^2 =m_1^2=m_2^2=m_3^2=m_4^2$ the model reduces to
the Lorentz-invariant Fierz-Pauli theory~\cite{Pauli:1939xp} (see also \cite{Arkani-Hamed:2002sp}).
Some other mass tunings have been discussed 
in~\cite{Arkani-Hamed:2003uy,Rubakov:2004eb,Dubovsky:2004sg}.
In particular \cite{Dubovsky:2004ud} (see also \cite{Gabadadze:2004iv})
describes a 
model with many similarities to our electrodynamical system --
there are massive transverse gravitational waves, and a seemingly instantaneous Newton's law.
We will not go into the details of the construction here, but merely comment that in this model it seems
the physical analysis we have done for the photon should apply: the gravitational potential will be 
slowly-instantaneous, producing
oscillations outside the light-cone with a characteristic frequency $m$. However, in massive gravity, one cannot
avoid strong coupling, because a Higgs mechanism is not known. Moreover, without an explicit
construction of gravitational vortices, it may not be possible to avoid an analog the Proca energy constraint
for a background gravitational field.

As a final word, it is perhaps worth briefly mentioning the sensitivity of our classical model to quantum corrections.
%The model is renormalizable, although technically one should add all the 
%additional Lorentz-violating operators. 
Naturally, standard model loops will contribute to additional relevant
Lorentz-violating operators. But coefficients of these terms
are at most around $m^2/m_e^2 \sim 10^{-34}$, which is well below even the strong bounds
$\sim 10^{-17}$ \cite{Lipa:2003mh} which already exist.

\begin{acknowledgments}
We thank S.~Dubovsky, T.~Gregoire, A.~Gruzinov, J.~D.~Jackson, R.~Rattazzi, 
and T.~Watari for helpful discussions.
G.D. is supported in part by David and Lucile
Packard Foundation Fellowship for Science and Engineering, and by NSF grant  PHY-0245068.

\end{acknowledgments}


\begin{thebibliography}{99}
\bibitem{footnote1}
%Essentially, the $m_0^2 A_0^2$ acts like a gauge-fixing term forcing us into temporal
%gauge $A_0=0$. 
For $m=0$,
%Although \eqref{scalpot} still holds, it is misleading. 
the Proca
constraint reads $\partial_t A_0=0$; this implies
that $A_0$ cannot mediate the electric force around
a charge, because it cannot move. 
Thus the Yukawa solution \eqref{scalpot} is not the physical one.
\end{thebibliography}
\end{document}